\documentclass[11pt,a4paper]{article}

\usepackage[utf8]{inputenc}
\usepackage[T1]{fontenc}
\usepackage{textcomp}
\usepackage{amsmath,amssymb}
\usepackage{graphicx}
\usepackage{booktabs}
\usepackage{array}
\usepackage{hyperref}
\usepackage[margin=1in]{geometry}
\usepackage{enumitem}
\usepackage{xcolor}
\usepackage{tikz}
\usetikzlibrary{arrows.meta,positioning,fit,backgrounds,calc}
\usepackage{fancyhdr}
\usepackage{titlesec}

\title{\textbf{Building Atomic Structures from Natural Language}}
\author{Pai Li\\[4pt]
\small State Key Laboratory of Materials for Integrated Circuits,\\
\small Shanghai Institute of Microsystem and Information Technology,\\
\small Chinese Academy of Sciences, Shanghai 200050\\[4pt]
\small Email: \texttt{lipai@mail.sim.ac.cn}}
\date{\today}

\begin{document}
\maketitle

\begin{abstract}
Large language models can write the input parameters of a materials simulation from a prompt, but they cannot build the atomic structure those parameters describe. A crystal, a surface slab or a mismatched interface must satisfy global constraints---periodicity along specific axes, a given vacuum thickness, a commensurate registry, sensible bond lengths. Generation one token at a time enforces none of them, and a violation is invisible in the output. We present ATLAS, a framework that makes structural construction scriptable. Seven components---analyzer, LLM skill, JSON format, build engine, validator, scorer and web interface---are organized around a component algebra of named structures and spaces. Every specification carries a check list, so a build is verified against the request rather than merely parsed from it, and builds are scored against the user's intent. Requests state physical sizes, and the engine computes the replication that reaches them. The translation is the one stochastic stage, and it is recorded, so a run replays with the model held fixed---which is what makes the framework's own behaviour measurable rather than merely observed. Fourteen structural classes are demonstrated, from bulk cells and nanoparticles through surfaces, defects and interfaces to amorphous networks and liquids. We outline how the same loop generates training sets for machine-learning force fields.
\end{abstract}

\section{Introduction}

Large language models can now write the input parameters of a materials simulation from a prompt. Agentic systems turn free-form requests into code-specific input files that run to completion on most benchmarks~\cite{genius}, and multi-agent frameworks orchestrate whole simulation pipelines from natural-language instructions~\cite{multiagent2025}. The structure the calculation runs on is a different matter. Atomic structures are geometric objects: atoms must sit at chemically sensible distances, periodic boundary conditions must hold, and each dimensionality---3D bulk, 2D slab, 1D wire, 0D cluster---carries its own vacuum and periodicity requirements. Multi-component systems need aligned interfaces with controlled gaps. Language models produce plausible-looking text that violates these constraints~\cite{llm_crystals,genms2024}, and the violation is invisible, because the coordinates are still well-formed numbers.

Fine-tuning does work for finite molecules, where the composition is fixed and the object is small enough to learn as a language~\cite{llm_geometry}. A periodic crystal, a surface or an interface is a different problem: its constraints are global. This paper asks how a software toolkit can let an AI comprehend and manipulate atomic structures, bridging natural language and atomic coordinates. ATLAS (Atomic Translation \& Language for Automated Structures) is our answer. One part of the system reads a structure and reports what is where, another turns a request into a formal specification, and a third builds the structure and checks that it answers the request.

Existing tools fall into three groups, and none of them can act as the structure builder in an LLM-driven workflow. Scripting libraries such as ASE~\cite{ASE}, pymatgen~\cite{pymatgen} and Atomsk~\cite{atomsk} are exact and reproducible, but every operation has to be written in Python, and the library has no idea which atoms lie on a surface. Graphical visualizers such as VESTA~\cite{VESTA} and OVITO~\cite{OVITO} show those spatial features immediately, but their operations are manual, leave no audit trail, and cannot be driven by a program. Language models accept the request but cannot produce the geometry. ATLAS takes one property from each group---spatial understanding from a visualizer, determinism from a library, the interface from a language model---and a JSON specification in the middle keeps the two jobs apart.

The comprehension half of the problem has received less attention. Crystal structure prototypes can be identified from a local atomic environment with a rotation-invariant descriptor~\cite{crystalproto}, and a mixed structure can be decomposed in 3D density space into chemical-formula components, each with a phase and a per-atom spatial role~\cite{cloudcomp}. What has been missing is less such analysis than its presentation to a language model as a stable, machine-readable document, and its use as the signal that verifies a build; C1 of this work is that interface (\S\ref{sec:c1}).

The contributions are: (1)~a comprehension layer that decomposes a structure into components with phases and per-atom roles and emits a single JSON document (C1); (2)~a component algebra in which every operation is a named component definition, so a build is a directed acyclic graph (C4); (3)~a specification format that separates what the model decides from what the engine executes, with named variables and a check list stating what the result must satisfy (C3); (4)~a validator that checks physical rationality under periodic boundary conditions, classifies dimensionality and repairs what it can justify (C5); (5)~a scorer that judges a build against the user's request rather than the specification, and awards nothing for what it cannot verify (C6); and (6)~an improvement scheme in which a curated test suite localizes a failure to a component (\S\ref{sec:optimization}).

\section{System Architecture}
\label{sec:architecture}

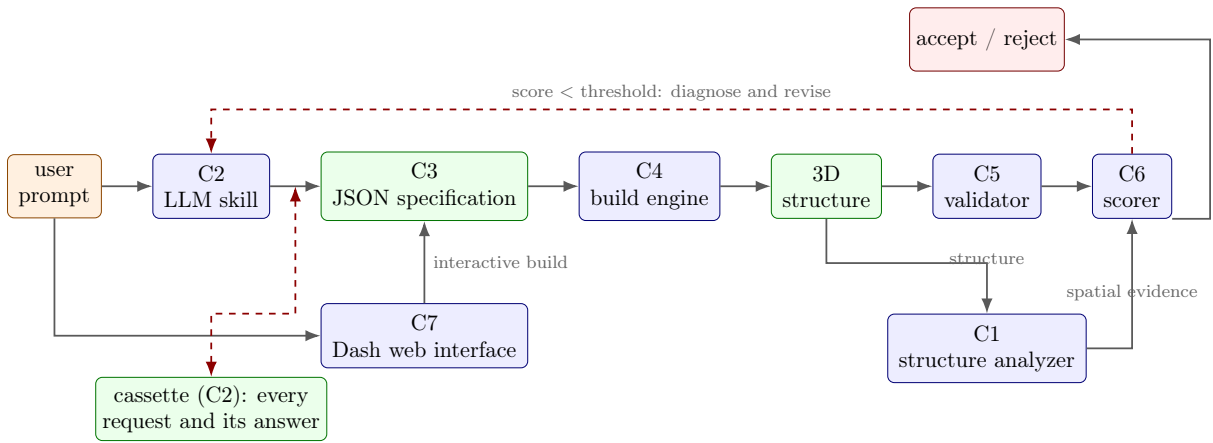
\begin{figure}[t]
\centering
\resizebox{\textwidth}{!}{%
\begin{tikzpicture}[
  font=\small,
  every node/.style={align=center},
  io/.style={draw, rounded corners=3pt, minimum height=1.0cm, inner sep=5pt,
             fill=orange!12, draw=orange!55!black},
  comp/.style={draw, rounded corners=3pt, minimum height=1.0cm, inner sep=5pt,
               fill=blue!7, draw=blue!45!black},
  art/.style={draw, rounded corners=3pt, minimum height=1.0cm, inner sep=5pt,
              fill=green!8, draw=green!45!black},
  ar/.style={-{Latex[length=2.4mm]}, thick, draw=black!65},
  fb/.style={-{Latex[length=2.4mm]}, thick, dashed, draw=red!55!black},
  lbl/.style={font=\scriptsize, text=black!60},
]
\node[io]  (prompt) {user\\prompt};
\node[comp, right=8mm of prompt] (c2) {C2\\LLM skill};
\node[art, right=8mm of c2] (c3) {C3\\JSON specification};
\node[comp, right=8mm of c3] (c4) {C4\\build engine};
\node[art, right=8mm of c4] (struct) {3D\\structure};
\node[comp, right=8mm of struct] (c5) {C5\\validator};
\node[comp, right=8mm of c5] (c6) {C6\\scorer};

\draw[ar] (prompt) -- (c2);
\draw[ar] (c2) -- (c3);
\draw[ar] (c3) -- (c4);
\draw[ar] (c4) -- (struct);
\draw[ar] (struct) -- (c5);
\draw[ar] (c5) -- (c6);

\node[comp, below=15mm of c5] (c1) {C1\\structure analyzer};
\draw[ar] (struct.south) -- ++(0,-7mm) -| (c1.north)
     node[lbl, pos=0.60, above] {structure};
\draw[ar] (c1.east) -- ++(6mm,0) -| (c6.south)
     node[lbl, pos=0.78, below] {spatial evidence};

\node[comp, below=13mm of c3] (c7) {C7\\Dash web interface};
\draw[ar] (c7) -- (c3) node[lbl, midway, right] {interactive build};
\draw[ar] (prompt.south) |- (c7.west);

\node[art, below=25mm of c2, inner sep=3pt, fill=green!8, draw=green!45!black]
     (cassette) {cassette (C2): every\\request and its answer};
\draw[fb, {Latex[length=2.4mm]}-{Latex[length=2.4mm]}]
     (cassette.north) -- ++(0,10mm) -| ($(c2.east)!0.5!(c3.west)$);

\draw[fb] (c6.north) -- ++(0,7mm) -| node[lbl, pos=0.25, above] {score $<$ threshold: diagnose and revise}
     (c2.north);
\node[art, above=13mm of c5, inner sep=3pt, fill=red!7, draw=red!45!black]
     (acc) {accept / reject};
\draw[ar] (c6.south east) -- ++(6mm,0) |- (acc.east);
\end{tikzpicture}}
\caption{The ATLAS pipeline. A user prompt is translated to a JSON specification by the LLM skill (C2~$\rightarrow$~C3); the build engine executes that specification into an atomic structure (C4); the validator checks and repairs physical rationality (C5); and the scorer evaluates the result against the \emph{original prompt} using spatial evidence from the structure analyzer (C6 with C1). A disagreement reported by the judgement feeds back to a diagnosis of which component failed, and the specification is revised. The translation is the one stochastic stage, and its answers are recorded: a run replays from the cassette with the model held fixed. The Dash interface (C7) exposes the same path interactively.}
\label{fig:architecture}
\end{figure}

ATLAS consists of seven interacting components organized in a pipeline with a feedback-driven optimization loop (Figure~\ref{fig:architecture}). The flow proceeds as follows. A user prompt is translated into a JSON specification by the LLM Skill (C2 $\rightarrow$ C3), and the Build Engine executes that specification into a structure (C4). The Validator then checks and corrects physical rationality (C5), and the Scorer evaluates the result against the original prompt, using spatial features from the Structure Analyzer (C6, with C1). If the judgement reports a disagreement, it is diagnosed and the JSON is revised for another iteration. The translation is the one stage that is not deterministic, so it is also the one stage that is recorded: every request and every answer is kept, and a run replays with the model held fixed (\S\ref{sec:optimization}), which is what lets everything below the translation be measured rather than merely observed.

\section{Structure Analyzer (C1)}
\label{sec:c1}

The analyzer answers one question for the rest of the system: \emph{where} in a structure is the region the user is talking about? Everything downstream depends on the answer: the scorer on whether the surface, interface or void the prompt asked for actually exists, the engine on acting on a region named in words rather than in coordinates.

It has two layers with a deliberate division of labour. The primary layer decomposes the structure in 3D density space and assigns every atom its spatial role from that segmentation alone; the complementary layer projects the structure along many directions and reports view-centric features. The split matters because the two answer different questions and fail differently: a role inferred from a projection is an artefact of the viewing direction, whereas a symmetry axis or an axial channel is something only a view can reveal. That document is what lets the analyzer serve as an interface rather than merely as a report.

\subsection{3D cloud-component analysis (primary)}

The primary layer is a 3D density decomposition. The partner \texttt{cloud\_comp} package~\cite{cloudcomp} segments a structure into \textbf{chemical-formula components}, classifies each component's \textbf{crystal phase} with a rotation-invariant prototype classifier~\cite{crystalproto}, and labels every atom with a spatial role --- bulk, surface, interface or vertex --- on the basis of that segmentation. The output is one machine-consumable JSON document with four top-level keys:

\begin{itemize}[leftmargin=*,itemsep=2pt]
    \item \texttt{model}: the formula, lattice, space group (via pymatgen's symmetry analysis, which uses spglib~\cite{spglib}) and per-component phase assignment;
    \item \texttt{labels}: one role per atom, in input order;
    \item \texttt{components}: for each chemical component, its atom indices, its own interior/boundary split and its periodicity;
    \item \texttt{summary}: aggregate counts (atom-class histogram, number of components, dominant symmetry).
\end{itemize}

Because the roles come from a density segmentation rather than from a per-atom geometric test, they remain meaningful for structures whose local environment is ambiguous --- a vertex atom on a faceted nanoparticle, an interfacial atom in a lattice-matched stack, an adatom above a surface. Those roles make referential spatial language executable: a request such as ``change all vertex atoms of the Pt cluster to Cu'' becomes a lookup in \texttt{labels} followed by an element substitution, with no hand-written geometric criterion.

The analyzer document is also the scorer's evidence base (\S\ref{sec:c6}): the scorer does not re-derive spatial facts from the structure when the analyzer has already derived them, and when the analyzer is unavailable the affected dimensions are reported as unverifiable rather than guessed.

One property of the decomposition decides whether it can be trusted at all: it must be a property of the \emph{crystal}, not of the cell we happened to write the crystal down in. Every cell the engine produces is a parallelepiped --- a slab cut along a Miller index, a lattice-matched interface, a re-oriented supercell --- and the coordinates that describe one naturally are fractional, not Cartesian. The density grid is therefore indexed in the cell's own basis: atoms are binned by their fractional position, with one code path for every cell shape and wrap-around periodicity that is exact in that basis. A rigid rotation of a crystal and its lattice vectors by the same orthogonal matrix --- which changes no bond length, no angle and no coordination shell --- therefore leaves the decomposition, and the formula read off it, unchanged; a sheared slab is identified as the material it is rather than as whatever its oriented cell happened to look like.
\subsection{The multi-view 2D projection lens (complementary)}

\begin{figure}[t]
\centering
\includegraphics[width=\textwidth]{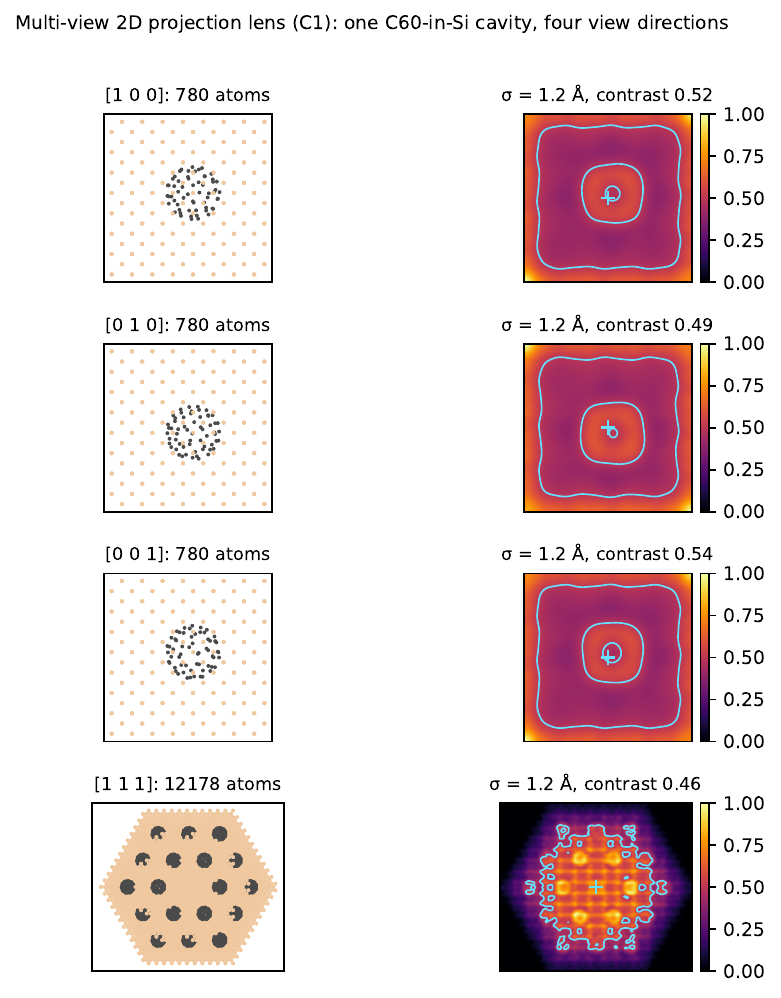}
\caption{The multi-view 2D projection lens (C1) applied to one structure, a C60 molecule inside a Si cavity. The same cell is projected along four directions; the density maps use a Gaussian kernel of $\sigma = 1.2$~\AA, wide enough that the host reads as connected and the cavity appears as a low-density core at the projected cell centre (marked $+$; the white contour is at half the maximum density). The quoted number is the host-to-cavity contrast measured along the centre line of that view, and the colour scale is the same in all four panels. The three axial views see the same cavity; the $\langle111\rangle$ view superimposes 12{,}178 atoms and resolves the host's own open channels instead, which is why a feature is only accepted when several views agree.}
\label{fig:projection}
\end{figure}

The second layer is explicitly a \emph{lens}, not a measurement instrument (Figure~\ref{fig:projection}). It projects the structure along a set of directions---crystallographic axes plus a Fibonacci-sphere spread for low-symmetry objects---and computes, for each view, a Gaussian density map~\cite{krone2012} and image-level features: symmetry from a Fourier transform in the spirit of 2D crystal image analysis~\cite{yang_crystal_fft}, void detection, and a convex hull. Agreement across views is used as a consensus signal: a symmetry axis or an axial void that appears in several projections is a property of the 3D object, whereas one that appears in a single view is an artefact of that projection.

The kernel width is a display parameter, not a measuring one. A small width resolves individual atomic columns but leaves the space between them at zero density, so a volume read off such a map would be an artefact of that choice rather than a property of the structure; a large width merges the structure into a featureless blob. Nothing in this work is measured from a density map: vacuum thickness and dimensionality are measured geometrically, from the atomic coordinates (\S\ref{sec:c5}).

Two further design rules keep this layer honest. First, \textbf{no per-atom descriptor is ever derived from 2D}: projected coordination numbers and bond lengths are artefacts of the projection, and the 2D layer is never permitted to assign an atom's role --- that is the 3D segmentation's job. Second, the lens is used for \emph{view-centric} questions (``does this structure have a channel along $c$?'', ``which axis is the symmetry axis?'') rather than for quantities that 3D measurement answers exactly. Table~\ref{tab:capabilities} lists the questions the two layers together can answer. The build path uses only the 3D half: the analyzer document is what the scorer verifies against and what the LLM skill is shown, whereas the two questions that only a projection can answer---symmetry axes and axial voids---are served by the interactive analysis views of \S\ref{sec:c7} and by the 2D selection operations, not by the model.

\begin{table}[h]
\centering
\caption{Spatial capabilities of the two analyzer layers. The rows marked ``2D lens'' are answered in the interactive analysis views; the rest are in the analyzer document the LLM skill is shown.}
\label{tab:capabilities}
\begin{tabular}{@{}l>{\raggedright\arraybackslash}p{9.4cm}@{}}
\toprule
\textbf{Capability} & \textbf{Question it answers} \\
\midrule
Component segmentation & Which chemical-formula components make up this structure, and how much of each is interior? \\
Phase classification & What crystal phase is each component in? \\
Per-atom roles & Which atoms are bulk, surface, interface, or vertex? \\
Dimensionality & Is this 3D bulk, a 2D slab, a 1D nanowire or a 0D cluster, and which axes are periodic? \\
Surface exposure & How many atoms are surface atoms, and with what coordination deficit? \\
Symmetry (3D) & What is the space group of the cell? \\
Symmetry axes (2D lens) & Along which directions does the projection show a symmetry axis? \\
Axial voids (2D lens) & Through which directions is there a continuous empty channel? \\
Interface extent & How many atoms lie at the boundary between two components? \\
\bottomrule
\end{tabular}
\end{table}

\section{The LLM Skill (C2)}
\label{sec:c2}

C2 is not a model. It is the specification of the translation task, kept in one skill document (\texttt{PROMPT\_TO\_JSON.md}, $\sim$680 lines) that is passed as context to any capable language model. Keeping the skill in a text file rather than in weights makes the contract versioned, reviewable and swappable without retraining, and the same document doubles as the human-readable specification of the JSON format.

The skill teaches four things. \textbf{The grammar}: the component algebra of \S\ref{sec:c3}---named components, the operations and their parameters, the two box-resize families, the space constructors, the named variables---together with the defaults to use when the prompt is silent (vacuum thickness, layer count, strain cap, gap width). \textbf{Worked specifications}: complete executable specifications for the recipes the algebra is for---a fullerene embedded in a host, a lattice-matched heterojunction, a core--shell particle, a liquid layer on a surface---and, for the remaining operations, a worked fragment or a table row: a vacancy, an antisite and an interstitial as site operations, a slab as the orient-then-cut sequence, a cluster and a supercell as the two ways to state a physical size. The complete ones are the specifications the test suite executes, not sketches. \textbf{The lattice discipline}: the hard part of the translation is not the vocabulary but the geometry, so the skill teaches the \emph{order} of operations---orient with \texttt{redefine\_lattice} before cutting with \texttt{slab}, match with \texttt{match\_lattice} before stacking with \texttt{translate above}---and states the strain cap within which an in-plane match counts as commensurate. \textbf{Self-checking}: the model attaches a \texttt{check\_list} to every specification it emits (\S\ref{sec:c5}), stating what the result must satisfy---no overlapping or isolated atoms, per-element-pair bond lengths, coordination ranges, vacuum along an axis, the target dimensionality and composition, the separation between components, clearance inside a carved space. That list is how the model states its intent in a form the validator can test rather than merely read.

The translation itself is a model step and there is no rule-based substitute: reading a request out of free text is the task a language model is for, and the interface reports that a model is needed rather than guessing at one. Everything below the translation is deterministic and runs with no model and no network, which is how the test suite and every figure here are produced: a specification, whether written by a model or by hand, is executed and checked by the same engine. The pipeline does not depend on any one model's reliability---the output is a declarative specification that is validated (C5) and scored against the prompt (C6), so a wrong translation is reported rather than silently executed. The boundary cuts both ways: because the translation is the one stage that is not reproducible, it is also the one stage that can be \emph{recorded}. Its answers are kept, and a run replays with the model held fixed (\S\ref{sec:optimization}), which is what turns the loop below it from a sequence of samples into an instrument.

\section{The JSON Specification and the Component Algebra (C3)}
\label{sec:c3}

C3 is the bridge language. A build is a \emph{graph} of named components --- the specification names the structures and spaces that take part and the operations that relate them --- rather than a flat list of modifications applied in order. A bridge language is needed because the two halves of the problem have incompatible notions of what a structure is: the language model manipulates \emph{names and intentions}, while the build engine manipulates \emph{coordinates and lattices}. The specification format is the contract between them, and its design goal is that every semantic decision the model makes is written down explicitly enough to be executed without further interpretation. Three ingredients carry that intent: named components the build can refer to, variables the engine resolves at execution time, and a check list stating what the finished structure must satisfy.

\subsection{Components: structure and space}

A specification is a graph of \textbf{named components}. Each component is one of two kinds:

\begin{itemize}[leftmargin=*,itemsep=2pt]
    \item a \textbf{structure} --- atoms together with a periodic box; or
    \item a \textbf{space} --- a region inside a periodic box, carrying no atoms.
\end{itemize}

A component is defined either by a \texttt{source} (a crystal from the structure library, matched fuzzily by formula) or by an \texttt{operation}. This yields the central design decision of the algebra: \textbf{every operation is itself a component definition}. There is no separate mutation step and no hidden state. An operation consumes one or two previously named components and emits a new named component, so the whole build is a declarative graph whose every edge is visible.

\subsection{Operations}

Operations are partitioned by arity and by the kinds they accept (Table~\ref{tab:operations}). The unary structure operations \texttt{supercell}, \texttt{resize\_box}, \texttt{redefine\_lattice}, \texttt{match\_lattice}, \texttt{slab}, \texttt{translate}, \texttt{rotate}, \texttt{change\_element} and \texttt{passivate} transform one structure. The space constructors \texttt{sphere}, \texttt{box}, \texttt{cylinder} and \texttt{ellipsoid} build regions. The three binary operations---\texttt{union}, \texttt{subtract} and \texttt{intersect}---are defined over every combination of kinds: structure with structure, structure with space, and space with space. Defining them over all combinations lets a cavity be carved, a shape cut out of a crystal, and two regions combined, all in one vocabulary.

The two box-resize families have different invariants, and conflating them produces wrong structures. Replication is the separate \texttt{supercell} operation. \texttt{resize\_box} has two families with different invariants: the \texttt{cartesian} family scales the box while holding Cartesian coordinates fixed (the correct choice when adding vacuum, and it warns when applied to a periodic axis, because that breaks periodicity), and the \texttt{fractional} family scales the box while holding fractional coordinates fixed (the correct choice when straining a lattice).

\begin{table}[h]
\centering
\caption{The component algebra: operations and the kinds they accept}
\label{tab:operations}
\begin{tabular}{@{}ll>{\raggedright\arraybackslash}p{7.3cm}@{}}
\toprule
\textbf{Arity} & \textbf{Operation} & \textbf{Effect} \\
\midrule
unary (structure) & \texttt{supercell} & replicate the cell along each axis \\
 & \texttt{resize\_box} & resize the box: \texttt{cartesian} or \texttt{fractional} family \\
 & \texttt{redefine\_lattice} & re-base the cell so a Miller index or direction becomes an axis \\
 & \texttt{match\_lattice} & find a commensurate in-plane cell and strain onto it \\
 & \texttt{slab} & cut $n$ atomic layers along the stacking axis and add vacuum \\
 & \texttt{translate} & move a component, or stack it \texttt{above} another with a gap \\
 & \texttt{rotate} & rotate the structure about an axis \\
 & \texttt{change\_element} & substitute one element for another, throughout the component \\
 & \texttt{remove} & delete the atom(s) at a named site: a vacancy, a divacancy, a cluster \\
 & \texttt{substitute} & replace the element at a named site: a dopant, an antisite \\
 & \texttt{insert} & put one atom at a named site: an interstitial \\
 & \texttt{passivate} & terminate dangling bonds with a chosen species \\
\midrule
space & \texttt{sphere}, \texttt{box} & construct a region inside a periodic box \\
 & \texttt{cylinder}, \texttt{ellipsoid} & \\
\midrule
binary & \texttt{union} & combine two components \\
 & \texttt{subtract} & remove one component from another \\
 & \texttt{intersect} & keep only the overlap \\
\bottomrule
\end{tabular}
\end{table}

\subsection{Variables and the check list}

Literal numbers are not always known when the specification is written --- the centre of a cavity depends on the host cell, and the height of a stack depends on the components plus the gaps between them. The algebra therefore provides \textbf{named variables} that the engine resolves at execution time. \texttt{box\_center}, \texttt{centroid} and \texttt{com} give positions; \texttt{surface\_site} and \texttt{interstitial\_site} (found by Voronoi analysis) give chemically meaningful sites; \texttt{literal} gives an explicit value; and \texttt{stack\_extent} sizes a final box from a list of components, the gaps between them and a vacuum margin. A variable may be referenced anywhere a geometric quantity is expected, so a specification stays valid when its inputs change.

Every specification may also carry a \textbf{\texttt{check\_list}}: a declarative statement of the physical properties the built structure must satisfy, listed in Table~\ref{tab:checks}. The check list is the interface between semantic intent and physical verification --- the model says what it meant, and the validator determines whether the engine delivered it.

Source matching is \emph{fuzzy}, so ``silicon'' resolves to the Si entry in the library without the model knowing file paths. And component references are by name, so a specification can be reordered, extended or partially replaced without renumbering anything.

Recognition is structural rather than nominal: a specification is V4 when its components carry an \texttt{operation}, or when it declares \texttt{variables} or a \texttt{check\_list}. Anything else is refused with the reason rather than guessed at, because a half-understood specification does not fail --- it builds a plausible but \emph{different} structure, a bare unit cell instead of the requested nanoparticle, and nothing in the result says so.

\section{The Build Engine (C4)}
\label{sec:c4}

The build engine turns a specification into an atomic structure---a \texttt{pymatgen} \texttt{Structure} object in the reference implementation. It evaluates the component graph as a directed acyclic graph: components are resolved in dependency order, each operation is applied to the components it names, and the result is bound to its own name. Because every operation is a component definition, the evaluation has no side effects on prior components --- a component can be referenced by several successors, and the graph can be partially rebuilt without recomputing or corrupting the rest.

\subsection{The lattice frame}

The engine's most important discipline is that geometry is expressed in the \emph{lattice frame}, never in global Cartesian coordinates. Slab cuts, stack translations and vacuum checks all work along a lattice vector or in fractional coordinates. Working in the lattice frame keeps oriented cells correct. The surface normal of a (110) slab, or the stacking axis of a lattice-matched interface, is generally not aligned with any global axis, and an operation that assumed it was would silently produce the wrong geometry.

Two mechanisms support this. Orientation is handled by \texttt{redefine\_lattice}, which re-bases a cell so that a requested Miller index (or a Cartesian direction) becomes a cell axis --- the prerequisite for cutting a slab with a specific surface index out of a primitive or conventional cell. Lattice matching is handled by \texttt{match\_lattice}, which searches for a commensurate in-plane supercell of two crystals, including the rotational registry between them, within a compression/stretching cap (5\% by default), strains the source onto the reference's in-plane lattice, and optionally registers the reference so that both sides of an interface share one matched box. Two planning helpers support the search: \texttt{best\_match\_direction} chooses the contact direction with the smallest resulting box when none is given, and \texttt{plan\_commensurate\_box} finds a single box that accommodates a whole stack of slabs within the strain cap.

Periodic-boundary handling is concentrated in a small set of helpers rather than scattered through the operations. A mean position is computed as a \emph{periodic} centroid, so a molecule straddling a cell corner does not report a centroid at the far corner. Unwrapping, likewise, is applied only along axes that contain a vacuum-sized gap: an axis spanned by the structure has no meaningful wrap, and unwrapping it would move a whole atomic plane by a lattice vector.

\subsection{A slab is cut between planes, and the request is about the material}

A thickness is a physical quantity the request names, and a slab can only be cut between atomic planes, so the sizes a cut can reach are the stacking repeat and its multiples. Rutile TiO$_2$(110) repeats every 3.25~\AA{}, so a request for ``about 8~\AA{}'' can be answered with a 6.5 or a 9.7~\AA{} cut but not with an 8~\AA{} one. Three rules decide what happens then. The quantity measured, ranked and reported is the \emph{material} thickness --- the span of the atoms the cut keeps, not the height of the cell they sit in --- and the two differ by the part-gap the cut leaves at each face, which is why ranking in cell height compares the request against a number the user never sees. A cut that would change the parent's element ratio is never taken over one that keeps it: rutile cut between repeats is Ti$_6$O$_{11}$, a different material from the one the request named, so a size no whole repeat can reach comes back thicker or thinner instead. And which of the two to prefer when they disagree is the request's to make: by default the cut is chosen for the fewest broken bonds, and a request that insists on the size takes the plane-bounded cut nearest to it, its surfaces unweighed. Every substitution is reported with the quantity asked for, the one built and the criterion that decided, so a size the build did not reach is stated in the log rather than inferred from the coordinates.

\subsection{Worked example: embedding a guest in a host}

The C60-in-Si build is seven operations, and it exercises most of the algebra:

\begin{enumerate}[label=\textbf{\arabic*}. ,leftmargin=*,itemsep=1pt]
    \item \texttt{supercell} --- expand the Si cell to $3\times3\times3$ to make room for the guest;
    \item a \textbf{sphere space} in the same box --- the cavity to be carved;
    \item \texttt{subtract} --- remove that space from the Si supercell, producing the host;
    \item \texttt{resize\_box} --- bring the C60 cell onto the host cell (the guest's own cell must be compatible before a union is meaningful);
    \item \texttt{translate} --- move the guest to the cavity centre, using a \texttt{box\_center} variable rather than a literal;
    \item \texttt{union} --- insert the guest into the host;
    \item the \texttt{check\_list} --- confirm that nothing overlaps, that the bond lengths are chemically sane, and that the result is still 3D bulk.
\end{enumerate}

Each step is a named component, so a failure anywhere in the build can be attributed to a named operation.

For molecular media the engine additionally supports a \textbf{molecule-aware} mode on the binary operations. A purely geometric cut removes atoms one at a time, and the atoms it removes need not belong to the same molecule: on a 60-molecule water box, an atom-wise cut retained 103 atoms in 41 bond-graph components, of which only 28 were intact molecules --- thirteen were fragments, hydrogens separated from their oxygens. With \texttt{whole\_molecules} set, the engine groups the structure into molecules by bond-graph connected components and decides membership once per molecule, so a molecule is kept or deleted whole (\texttt{whole}, \texttt{centroid} or \texttt{any}). The same cut then retains 84 atoms as 28 intact molecules and nothing else. Molecular liquids are generated separately by random sequential insertion of rigid molecules at a target density, with every insertion checked against a minimum-image contact floor derived from the same covalent-radius criterion as the bond graph, so that two distinct molecules can never be mistaken for one bonded molecule.

\section{The Validator (C5)}
\label{sec:c5}

The validator decides whether a built structure is physically rational, and it is deliberately separate from the engine that built it: a build is judged against the specification's own \texttt{check\_list}, not against the engine's assumptions. That separation is what turns the specification from a description into a claim that can fail.

Two mechanisms do the work. A set of geometric checks---overlap, isolation, bond length, coordination, vacuum, dimensionality, composition, separation, cavity clearance---answers whether the structure is chemically and geometrically sound. When a check fails, a bounded auto-fix loop attempts the repair the specification justifies and reports what it changed rather than silently succeeding.

\subsection{The check list}

\begin{table}[h]
\centering
\caption{Physical checks available to a specification}
\label{tab:checks}
\begin{tabular}{@{}l>{\raggedright\arraybackslash}p{10.0cm}@{}}
\toprule
\textbf{Check} & \textbf{What it asserts} \\
\midrule
\texttt{no\_overlap} & No two atoms closer than a tolerance --- the primary defence against incoherent coordinates. \\
\texttt{no\_isolated} & No atom left without a neighbour, which is how a broken fragment or a stray atom shows up. \\
\texttt{bond\_length} & Every bonded pair lies within a range derived from covalent radii; defaults per element pair, overridable. \\
\texttt{coordination} & Per-element coordination numbers fall inside specified ranges; where a specification states none, the expectation is read from the phase the analyzer identifies, measured on that phase's own interior, and with nothing identified the check reports the observed distribution rather than inventing bounds. \\
\texttt{vacuum\_layers} & Vacuum is (or is not) present along a given axis, with a minimum thickness. \\
\texttt{dimensionality} & The measured dimensionality matches the target ($\text{0D}$--$\text{3D}$). \\
\texttt{composition} & The elemental composition matches the requested one. \\
\texttt{separation} & Two named components are separated by at least a given distance --- the interface-gap check. \\
\texttt{no\_atoms\_in\_space} & A named region is empty, optionally restricted to one element --- the cavity-clearance check. \\
\bottomrule
\end{tabular}
\end{table}

All geometric checks are periodic-boundary aware: distances are computed against periodic images, and a region test uses fractional coordinates, so a structure is not reported as invalid merely because an atom sits on a cell face.

\subsection{Dimensional classification}

\begin{figure}[t]
\centering
\includegraphics[width=\textwidth]{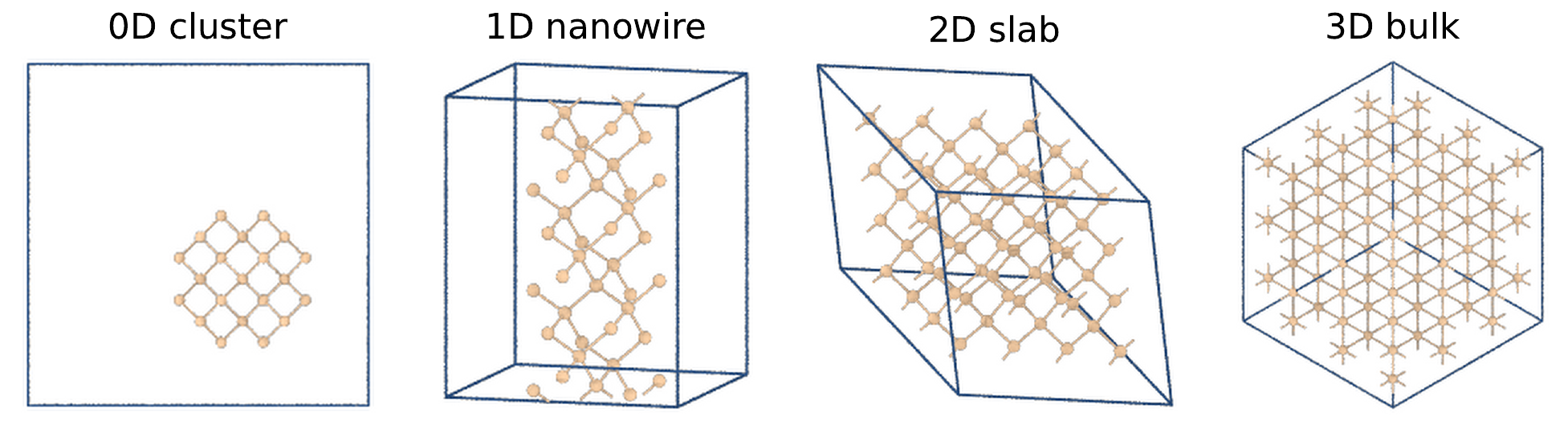}
\caption{Dimensional classification of the same material in four geometries, rendered with OVITO. A single shared predicate decides all four cases---a cell axis counts as periodic when no vacuum-sized gap separates the atoms along it---and the cell frame is drawn so that the periodic axes are visible.}
\label{fig:dimensionality}
\end{figure}

Dimensionality is the property most often stated in a prompt (``a slab'', ``a nanoparticle'') and the easiest to get wrong, so it is measured from the structure's own cell rather than inferred from how it was built (Figure~\ref{fig:dimensionality}). The validator decides, for each cell axis independently, whether that axis is periodic, using a single shared predicate: an axis is periodic when it contains no vacuum-sized gap. The threshold is an absolute distance in \AA ngstr\"oms, not a fraction of the cell: one predicate can then serve the validator, the scorer and the cross-view consensus, so the same axis is periodic to every component that asks. The number of periodic axes then gives the classification: none is a 0D cluster, one a 1D nanowire, two a 2D slab, three 3D bulk.

The subtlety the shared predicate resolves is that a structure's dimensionality is a property of its \emph{cell}, not of its components. The interior of a nanoparticle is bulk-like, so a component-level analysis reports it as three-dimensional even though the object is a cluster; measuring the cell avoids that error.

\subsection{Iterative auto-fix}

When a check fails, the validator can attempt a repair rather than only reporting the failure. Auto-fix is explicitly bounded: it corrects what it can justify from the specification --- enlarging a cell whose vacuum is too thin, for instance --- and it must terminate. Two conditions enforce that: the repair target is expressed so that it is actually reachable, and a repair that would change a genuinely three-dimensional cell is not attempted at all. The loop stops when the cell stops changing instead of re-applying the same ineffective correction.

Required checks propagate. A build whose mandatory checks fail is reported as a failure to the caller, with the failing checks named, rather than being logged as a success with a note attached; the advisory behaviour is retained only for direct callers that explicitly ask for it.

\section{The Scorer (C6)}
\label{sec:c6}

A build can satisfy every structural check and still not be the object the user asked for. The scorer closes that gap: it judges the built structure against the original prompt, never against the specification derived from it, so an error made during translation is detected rather than inherited.

It is a structure parser, a material identification, a language model and the prompt. The parser reads the structure and states facts (Table~\ref{tab:parser}) --- formula and per-element counts, the cell, the largest empty gap along each cell axis, the measured dimensionality, the largest distance between any two atoms, and, per component, the material the analyzer takes it to be. The model is given those facts beside the original request and answers one question: \emph{is this the object the request describes?} It reports the disagreements it can establish, each with an aspect and a severity (Table~\ref{tab:judge}).

The material identity is not left to that model. A formula does not name a substance: rutile, anatase and brookite are all TiO$_2$, and diamond and graphite are both C. The analyzer already answers what a component is --- its density decomposition gives the region and its formula, and its local-environment classifier gives the structure type, one of 272 prototype classes or ``amorphous'' --- and when a request names a phase, the identified phase is compared with it directly. A contradiction is a \emph{material} disagreement the scorer applies itself, at the same severity as one the model reports, and the two are not counted twice. The same identification supplies the expected coordination when a specification does not state it (\S\ref{sec:c5}).

The judgement is deliberately not an extraction. A schema would have to anticipate every way a request can be phrased, and a fact the schema does not carry is a defect it cannot see. An earlier revision of this component did extract: it read elements, dimensionality and a few counts out of the prompt with rules and scored five weighted dimensions. Measured against the builds in \S\ref{sec:demo} it passed a four-atom cell as a fourteen-\AA{}ngstr\"om nanoparticle at 85/100, a C60 host containing thirty carbons at 95/100, and an oxygen-deficient TiO$_2$ slab at 85/100. Asking instead whether the two agree needs no such anticipation, and it found all three.

The physics of the structure is not the model's business. The specification's own check list and the geometric checks run on the structure directly, and a build that answers the request but fails them does not pass, so a generous verdict cannot carry a broken structure. Nor is there a fallback score: a build that no model has judged is reported as \emph{unjudged}, because a number produced by rules that cannot read the request is worse than an absent number.

\begin{table}[h]
\centering
\caption{What the parser reports to the scorer. These are the facts the model is given; it never sees the specification. The identified materials are per component, with the formula the density decomposition gives it and the structure type the local-environment classifier gives it.}
\label{tab:parser}
\begin{tabular}{@{}l>{\raggedright\arraybackslash}p{8.6cm}@{}}
\toprule
\textbf{Fact} & \textbf{Example} \\
\midrule
formula and per-element counts & \texttt{Si}$_{465}$\texttt{C}$_{60}$, C 60, Si 465 \\
number of atoms & 525 \\
cell dimensions & 21.7 $\times$ 21.7 $\times$ 21.7 \AA{} \\
measured dimensionality & 3D bulk, from the structure's own cell \\
largest distance between any two atoms & 12.90 \AA{} --- the size of the object \\
largest empty gap along each cell axis & 4.08, 4.08, 4.08 \AA{} --- the vacuum \\
identified materials & \texttt{Si}: diamond (A4), crystalline; \texttt{C}: not resolvable --- formula and structure type per component \\
\bottomrule
\end{tabular}
\end{table}

\begin{table}[h]
\centering
\caption{What the judge reports. Each disagreement carries an aspect, what was requested, what was built, and a severity; a \emph{major} disagreement fails the build. The score is $100 - 25\,n_\text{major} - 8\,n_\text{minor}$.}
\label{tab:judge}
\begin{tabular}{@{}l>{\raggedright\arraybackslash}p{9.4cm}@{}}
\toprule
\textbf{Aspect} & \textbf{Reported when} \\
\midrule
material & a requested element or compound is absent, or the structure is a different material \\
stoichiometry & the request states an atom count or a ratio and the structure does not match it \\
size & the request states how large the object should be and it is off by more than about a third \\
dimensionality & the requested dimensionality is not the one measured \\
vacuum & a vacuum gap is asked for, or asked to be absent, and it is not \\
shape & the request names a shape --- spherical, Wulff, core--shell, wire --- that the structure does not have \\
\bottomrule
\end{tabular}
\end{table}

\section{The Web Interface (C7)}
\label{sec:c7}

C7 is the interactive entry point: a Dash application~\cite{dash} that exposes the same pipeline in a browser, for the case where a user wants to inspect and steer a build rather than script it. The web layer is not a separate implementation: it calls the same build engine, validator and scorer described above, so an interactive build and a headless build of the same specification produce the same structure.

\begin{figure}[t]
\centering
\includegraphics[width=\textwidth]{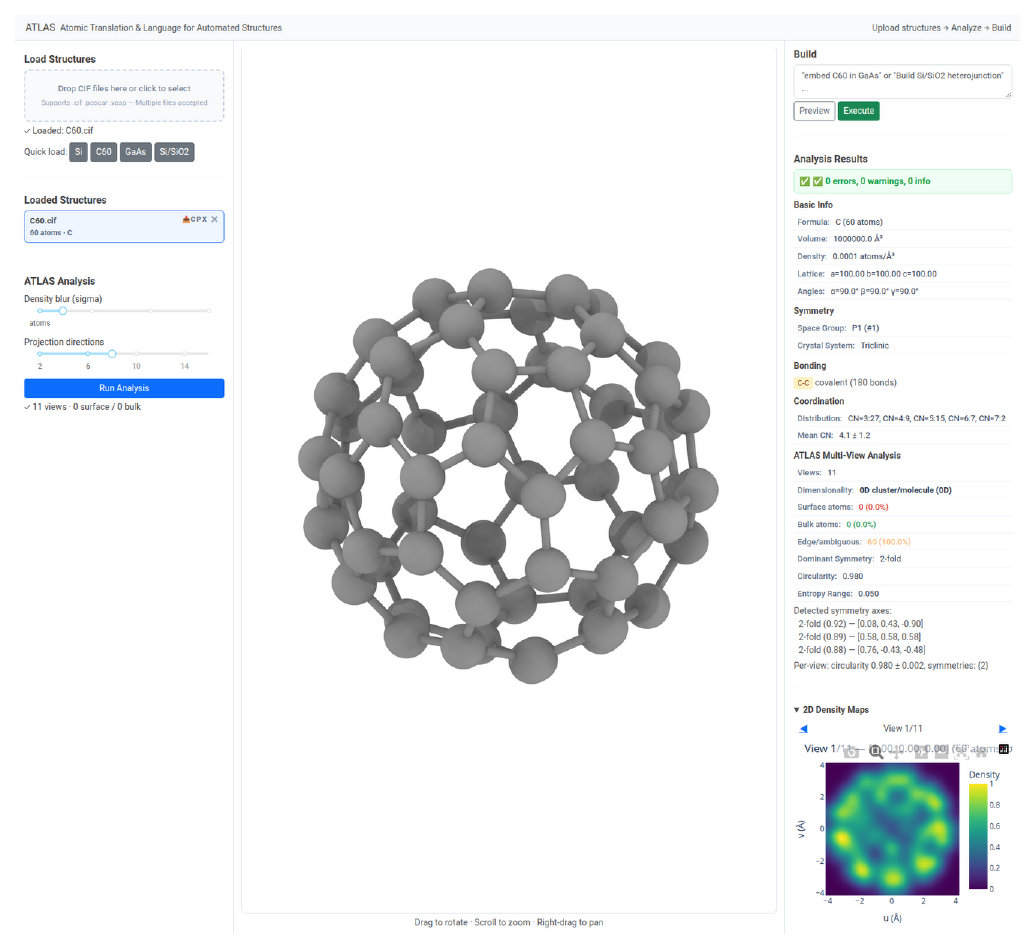}
\caption{The web interface (C7), shown after loading a C60 structure and running the analysis. Left: loading controls, the loaded structures, and the analysis controls --- the kernel width and the number of projection directions are exposed because they are choices a user may want to vary, and the same controls exist as parameters on the headless path. Centre: the loaded structure in the 3D viewer. Right: the build panel and the analyzer's output, which reports the dimensionality, the per-atom role fractions, the dominant symmetry with its detected axes, and the coordination and bonding statistics, above the 2D density maps that can be stepped through one projection direction at a time. The web layer is not a second implementation: every action here resolves to the specification the build engine and validator already consume.}
\label{fig:web}
\end{figure}

Figure~\ref{fig:web} shows the interface; it provides three things. The build panel takes model credentials --- an endpoint, a model name and an API key --- because the translation step needs a model and a build cannot proceed without one. \textbf{Analysis views} expose the comprehension layer directly: multi-view 2D projections with their density maps, per-atom role colouring from the analyzer document, and symmetry and void summaries. \textbf{A build panel} takes a prompt, shows the JSON specification that the LLM skill produces for it, and lets the user inspect or edit that specification before executing it --- the specification is visible before it is executed. \textbf{An interactive 3D viewer}, built on Crystal Toolkit~\cite{crystaltoolkit}, renders the executed structure and the cell so that the result can be rotated and checked by eye. Structures built during a session are held in a browser-side collection, so several candidates can be compared and the one that should be carried forward selected.

Two properties of the web layer are architectural rather than cosmetic. Every interactive action resolves to a JSON specification, which can be exported and re-executed headlessly --- so anything done by clicking is reproducible as text.

\section{Automatic Improvement Scheme}
\label{sec:optimization}

The framework is meant to be improved by measurement rather than by inspection, in four steps.

\textbf{A curated assertion suite.} Tests with \emph{known spatial answers}---an integer-volume re-oriented cell, a slab whose span is exactly one quarter of the lattice constant, an in-plane registry angle inside a stated range, a structure that must return \texttt{None}---turn vague expectations into pass/fail assertions. Because the answers are known independently of the implementation, the suite catches a change that makes the code self-consistent but wrong. It is 191 assertions over sixteen files, and the ones that carry the most weight are the ones whose answer could be written down before the code existed: the first-shell coordination of twenty-five textbook crystals and five molecules, each a number no implementation is free to disagree with; the reported formula and structure type of library crystals, which must survive a rigid rotation; the geometric identities above. The rest pin behaviour that has no closed form but does have a contract---the schema a specification may carry, the components the engine may consume, the guarantees the cassette makes about never inventing an answer and never recording a rejected one.

\textbf{Component-attributable scoring.} Each build is reported per dimension (\S\ref{sec:c6}) and each specification names the components it uses, so a failure localizes to a dimension and the dimension to a component: a build whose dimensionality dimension fails while its composition dimension passes is an orientation or vacuum problem in the lattice operations, not a material-resolution problem. That is what keeps the loop productive instead of a search over the whole codebase.

\textbf{The translation as a recorded input.} Three of the four stages are deterministic; the first is not, and one stochastic stage was enough to make the whole pipeline unmeasurable. Two runs of the same code differed for two reasons at once---the change under test and the sampling---so a one-case difference between them could be attributed to neither, and a defect that appeared in one run and not the next could not be reproduced at all. The remedy is not to make the model deterministic, which it is not, but to treat its answers the way the engine treats its inputs: every request and the answer it drew are written to a \emph{cassette}, and a later run is answered from the cassette. An unchanged pipeline is answered exactly, and replays offline with no key and no rate limit: the recorded run took twenty-six minutes with the model live and about twenty to replay, because the engine and the checks still run and those are what is being measured. A pipeline whose build has changed is answered from the position that call occupied in the recorded run (same kind of call, same case, same round), which holds the model fixed while the code moves. The second is a counterfactual rather than a repetition, so the two are counted apart and the count travels with the run: a run with no positional falls reproduces the earlier one answer for answer, and a run with some is a comparison rather than a repetition. That count is what tells a reader which of the two they are looking at.

\textbf{Diagnose, fix, re-measure.} A failure pattern becomes a targeted change, and the suite is re-run to confirm that the target improved and that nothing else regressed.

The four steps together are what turn the framework's own behaviour into a measurement. A specification is executed and judged identically whether a model wrote it or a person did, so the engine can be tested against its own specification rather than against a transcription of it. A cassette makes the translation an input, so a change is measured against a fixed set of translations rather than against a fresh sample. And the per-dimension breakdown reports a change against the component it touched, so a result is attributable rather than merely observed. The demonstration that follows is the outcome.

\section{Demonstration}
\label{sec:demo}

Every structure in this section was produced end-to-end by the component algebra from a specification, with no hand-placed coordinates, and every figure is regenerated from real pipeline output by a single script. The figure specifications live in that script, so the figures replay from a fixed specification with no model in the loop at all. The translation is demonstrated separately: the same fourteen prompts are put to the skill of \S\ref{sec:c2}, and the judgement of each attempt is fed back to the model for a revision, with the repair loop allowed four rounds.

That run is recorded, and it is reproducible in a stronger sense than a published number usually is. Every request the model was asked and every answer it gave are in the repository as a cassette (\S\ref{sec:optimization}); replaying it re-runs the whole demonstration with no model, no key and no network, and returns the same fourteen outcomes, round by round. The translation is the one stage that is not deterministic, and the cassette is what takes it out of the comparison.

In the recorded run thirteen of the fourteen closed, eleven of them on the first attempt, and the repair loop closed the other two within the four rounds it allows. The one it could not close is the informative one: the Si(111)$|$GaAs(111) request came back four times as a single intermixed Ga--Si--As slab rather than two stacked slabs, and the judgement said so each time. That is the framework's claim rather than a counterexample to it --- the wrong translation was reported, with the aspect it got wrong, not silently built. How many need the loop at all is not a property of the code alone: across the runs of these fourteen prompts archived in the repository the number solved on the first attempt has ranged from eight to fourteen, and so does the structure a case comes back as --- the C60-in-Si cell has been a 525-atom cell in one run and a 2762-atom cell in another, both answering the request.

Checking a build takes both halves, and neither substitutes for the other. The judgement reads the request, which no schema can fully anticipate: a stated size, a named facet, a stoichiometry, a shape. The check list measures the result, which a judgement can wave through: it is the structure, not the sentence, that says whether the vacuum is there and whether an atom was stranded. A failure in either is reported per aspect, and the score is reported per dimension (\S\ref{sec:c6}), so a disagreement names what it is about. Every build below is judged against its own request, and the judgement is what the table reports.

\begin{figure}[t]
\centering
\includegraphics[width=\textwidth]{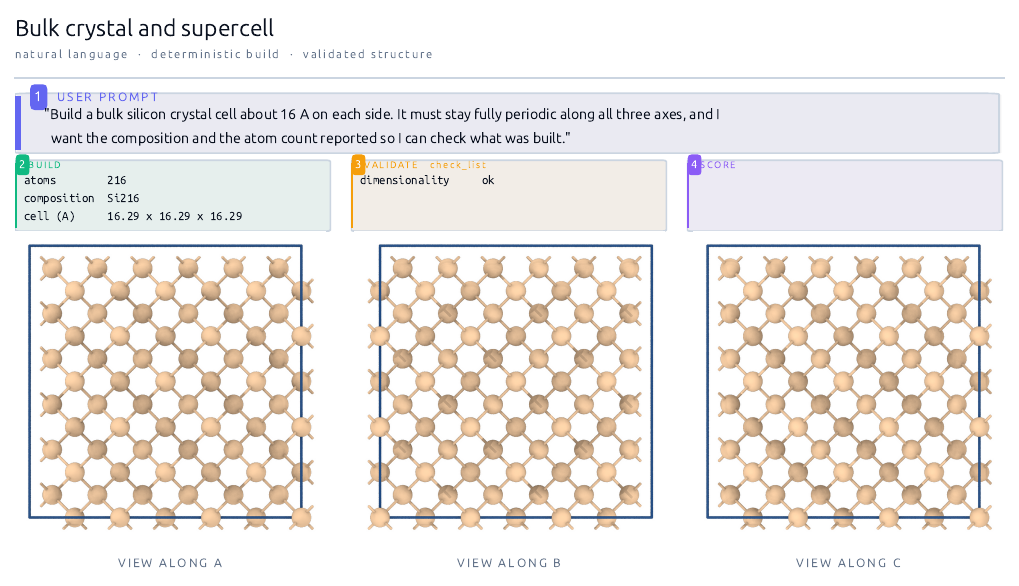}
\caption{Bulk crystal. The request states a size ("about 16~\AA{} on each side"), not a replication; the engine chooses the $3\times3\times3$ cell that reaches it. The box is the periodic cell, and 216 atoms are drawn in it.}
\label{fig:case-bulk}
\end{figure}

\begin{figure}[t]
\centering
\includegraphics[width=\textwidth]{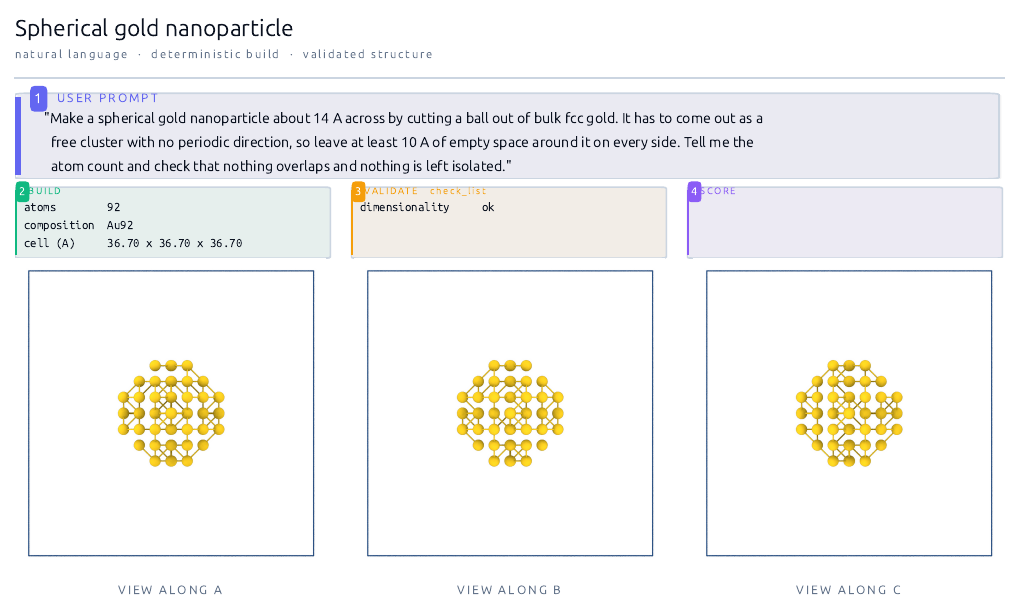}
\caption{0D nanoparticle by sphere intersection. A ball is cut from bulk fcc gold; the result has no periodic axis, so the cell is the vacuum around the cluster and every view shows it.}
\label{fig:case-nanoparticle}
\end{figure}

\begin{figure}[t]
\centering
\includegraphics[width=\textwidth]{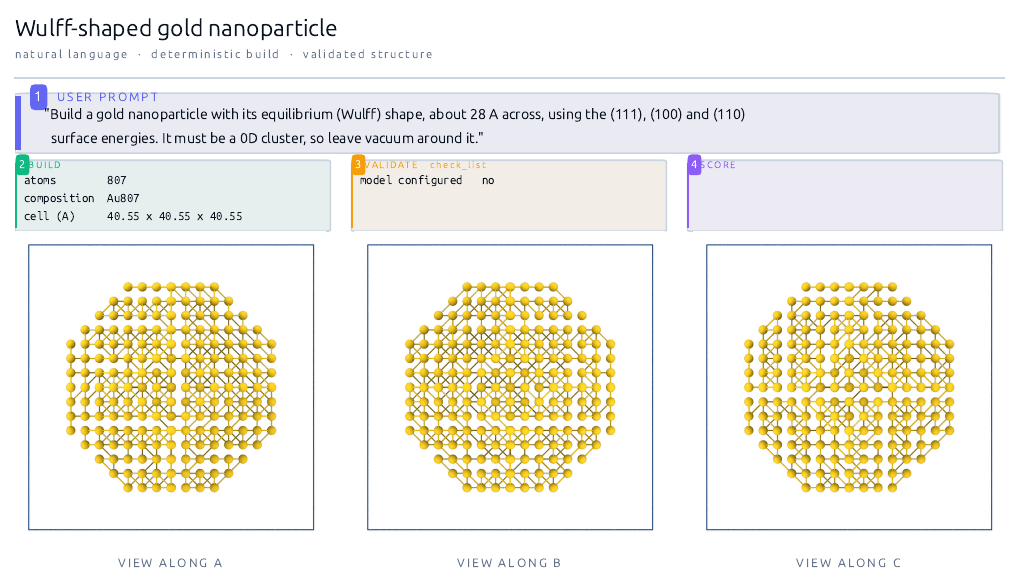}
\caption{Equilibrium crystal shape. The faceted particle is the Wulff construction from the (111), (100) and (110) surface energies, built as the intersection of the corresponding half spaces.}
\label{fig:case-wulff}
\end{figure}

\begin{figure}[t]
\centering
\includegraphics[width=\textwidth]{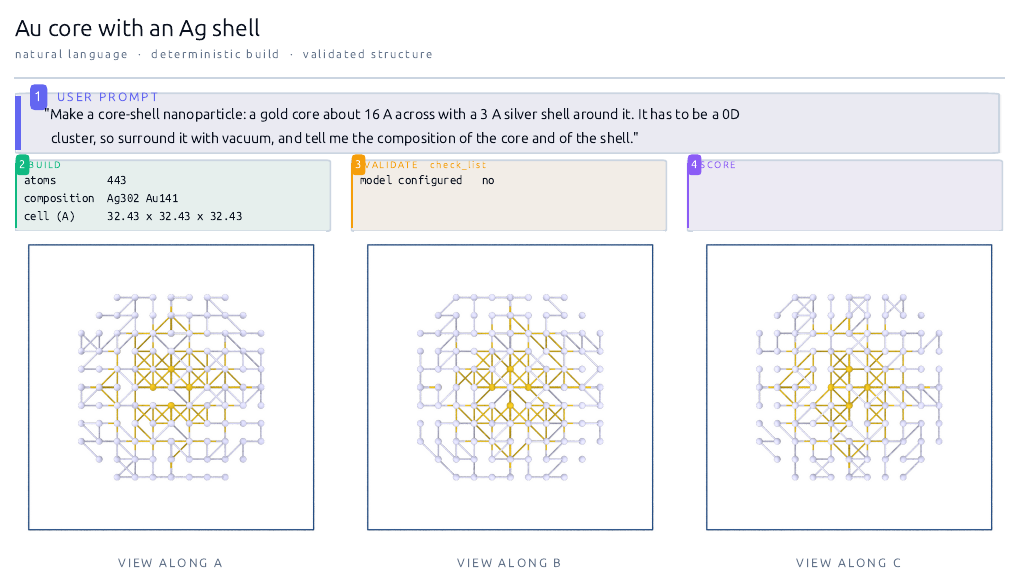}
\caption{Core--shell nanoparticle: a 16~\AA{} Au core inside a 3~\AA{} Ag shell. The core is drawn with a small ball radius so it shows between the shell atoms, and no cut-away is used. A clean separation --- every Au inside every Ag --- is what the case checks.}
\label{fig:case-core-shell}
\end{figure}

\begin{figure}[t]
\centering
\includegraphics[width=\textwidth]{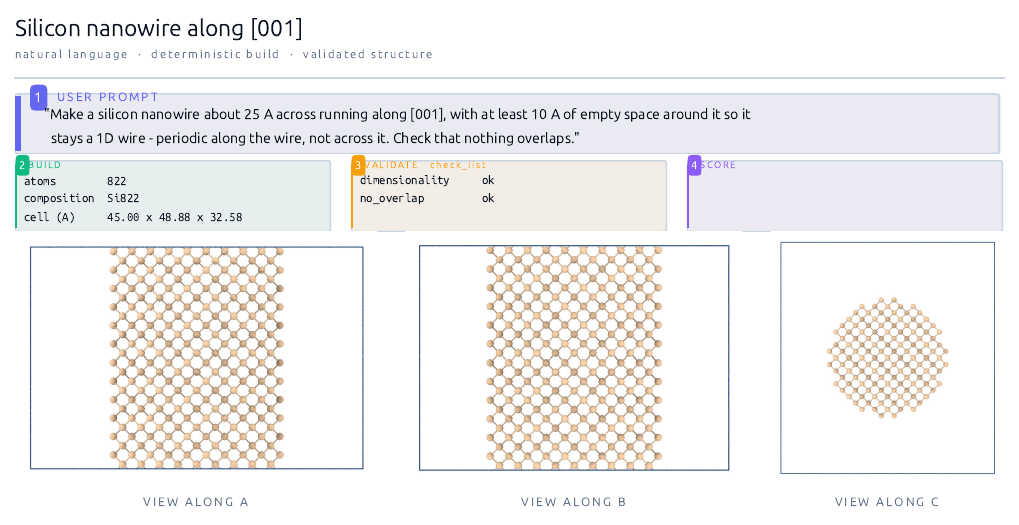}
\caption{1D nanowire. A cylinder is intersected with a large block, giving a wire periodic along its axis and surrounded by vacuum; the view along the axis is the circular cross-section.}
\label{fig:case-nanowire}
\end{figure}

\begin{figure}[t]
\centering
\includegraphics[width=\textwidth]{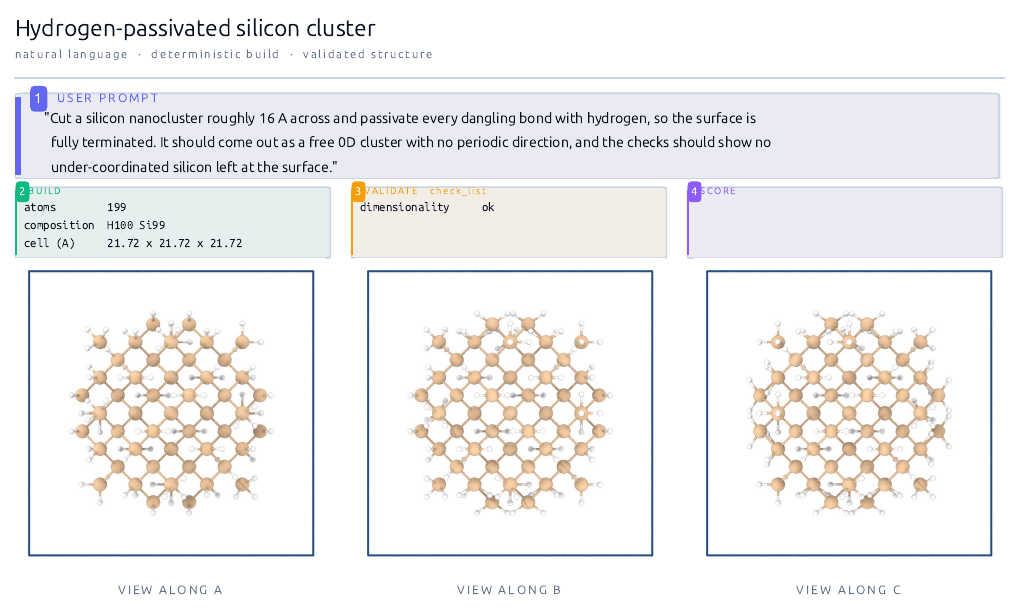}
\caption{Passivated nanocluster. Every dangling bond of a Si cluster is terminated with hydrogen, so the surface is chemically closed rather than left with unsaturated Si.}
\label{fig:case-passivated}
\end{figure}

\begin{figure}[t]
\centering
\includegraphics[width=\textwidth]{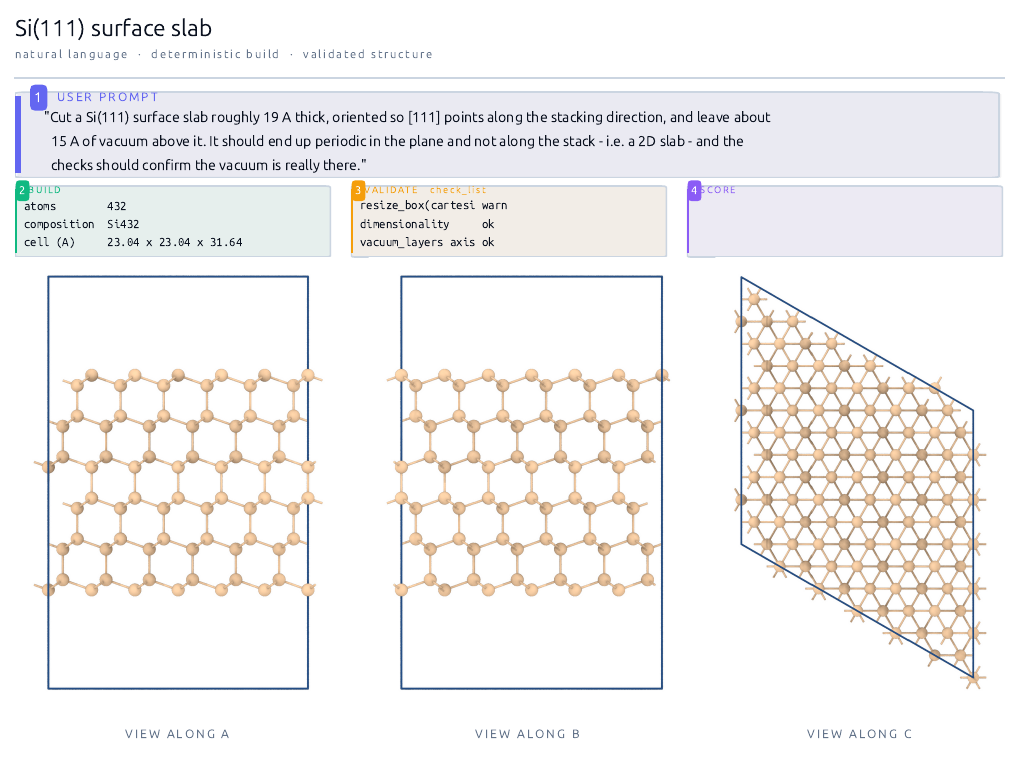}
\caption{2D surface slab. A Si(111) slab six bilayers thick with vacuum above it: the request gives a thickness and a vacuum in \AA ngstr\"oms, and the box shows both, so the vacuum the request asked for can be checked against what was built. The cut is placed between bilayers, so each surface plane keeps three bonds rather than one.}
\label{fig:case-slab}
\end{figure}

\begin{figure}[t]
\centering
\includegraphics[width=\textwidth]{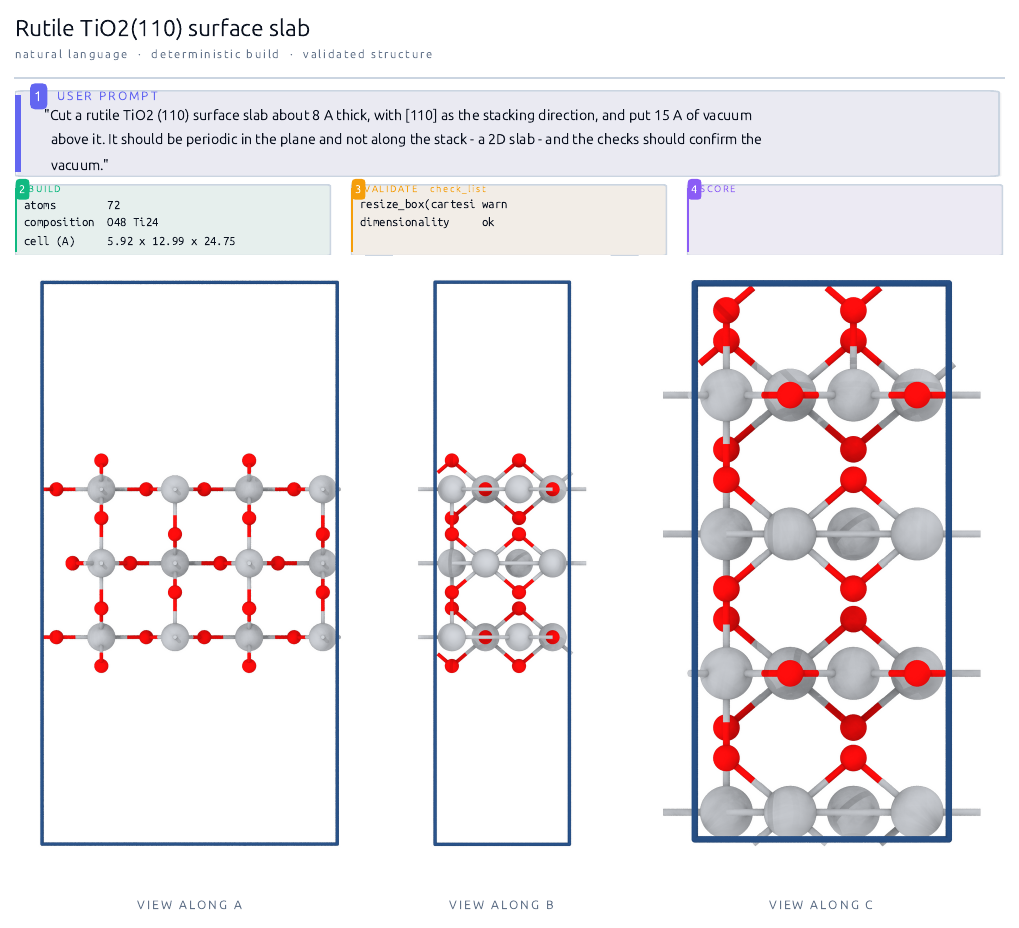}
\caption{Oxide surface. The same construction for a rutile TiO$_2$(110) slab, where the request names the polymorph and the facet but still gives the thickness and vacuum as lengths.}
\label{fig:case-tio2}
\end{figure}

\begin{figure}[t]
\centering
\includegraphics[width=\textwidth]{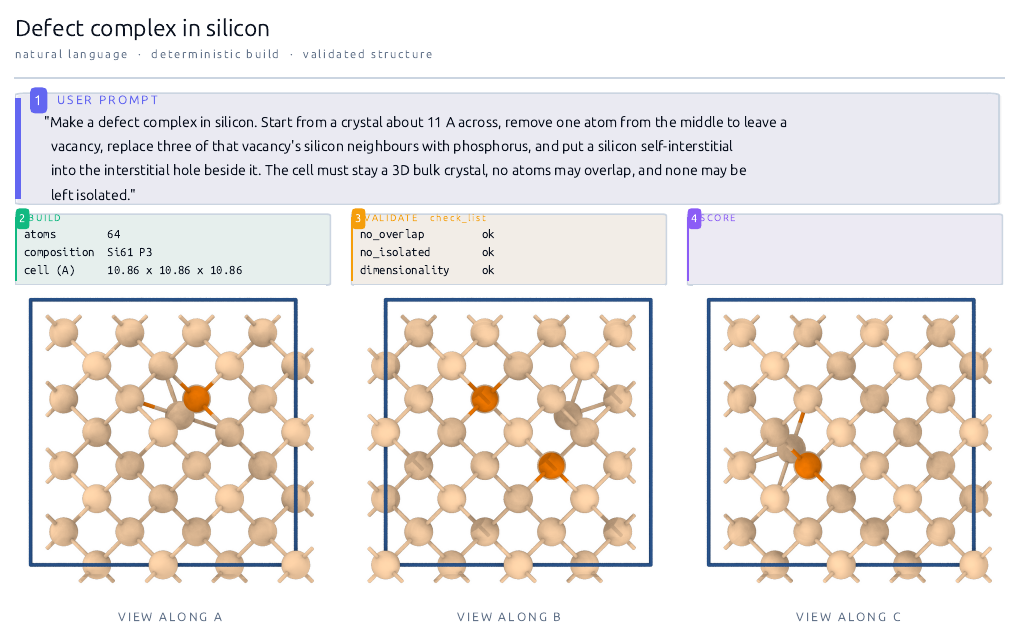}
\caption{Point defects. An intrinsic defect decorated by impurities: a vacancy, three phosphorus atoms on the neighbours it left behind, and a self-interstitial in the hole beside it. Three operations on named sites, one cell, and the three checks confirm the crystal is still bulk, still free of overlaps, and has not stranded a neighbour of the vacancy.}
\label{fig:case-defect}
\end{figure}

\begin{figure}[t]
\centering
\includegraphics[width=\textwidth]{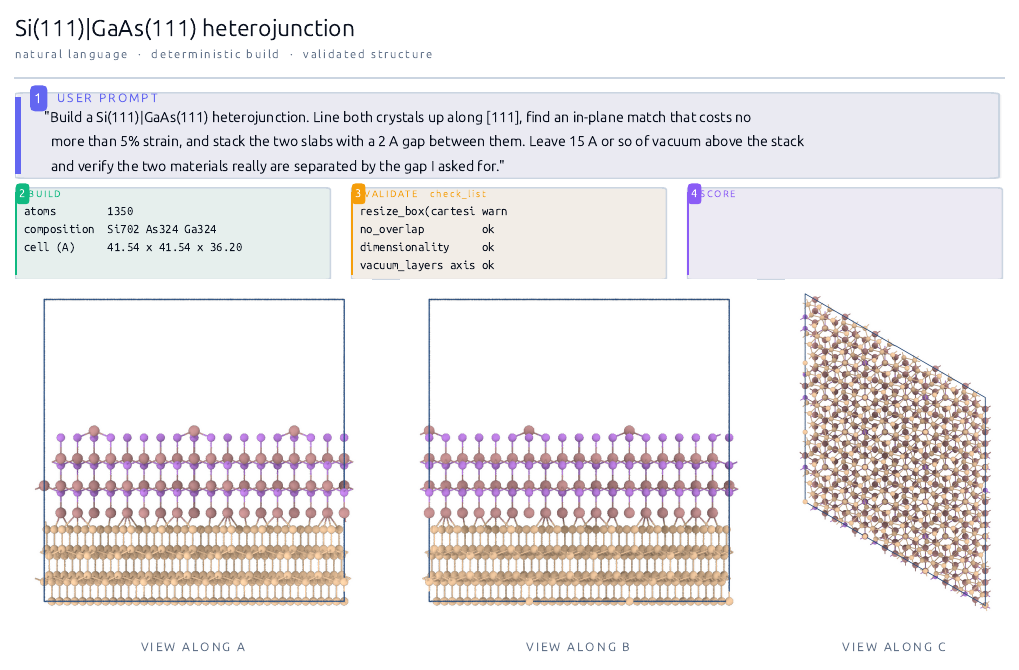}
\caption{Lattice-matched interface. Both crystals are oriented along [111], a commensurate in-plane cell is found within a 5\% strain cap, and the slabs are stacked with a 2~\AA{} gap. The lateral views show the layer order and the gap.}
\label{fig:case-hetero}
\end{figure}

\begin{figure}[t]
\centering
\includegraphics[width=\textwidth]{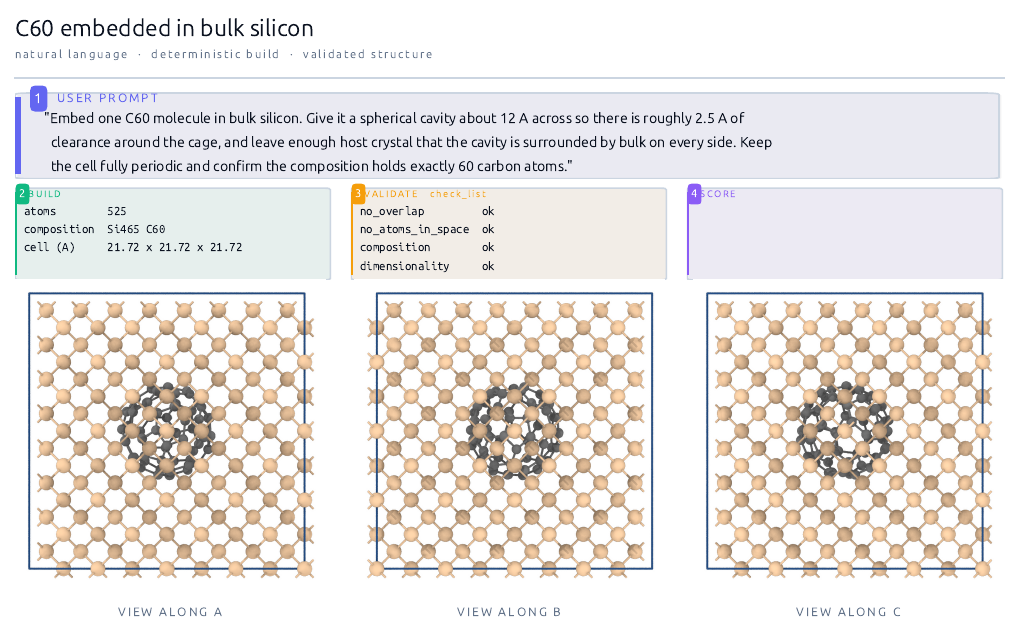}
\caption{Guest embedded in a bulk host. A cavity is carved out of bulk silicon and the fullerene placed in it. The guest is visible along the cell axes because the cavity lies in the line of sight --- no cut-away is needed.}
\label{fig:case-c60}
\end{figure}

\begin{figure}[t]
\centering
\includegraphics[width=\textwidth]{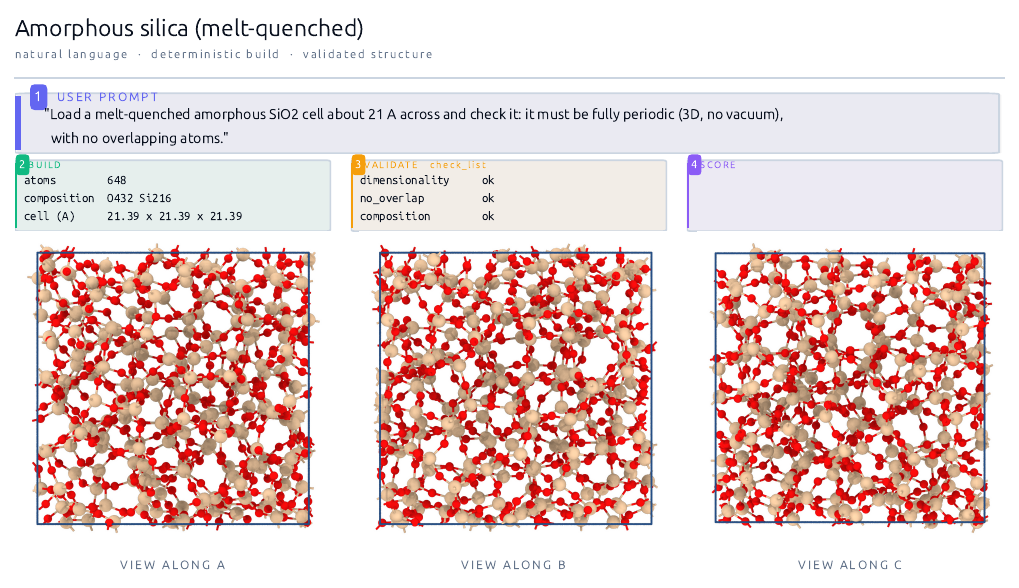}
\caption{Amorphous solid. A melt-quenched SiO$_2$ cell is loaded and checked: fully periodic, no vacuum, no overlapping atoms.}
\label{fig:case-amorphous}
\end{figure}

\begin{figure}[t]
\centering
\includegraphics[width=\textwidth]{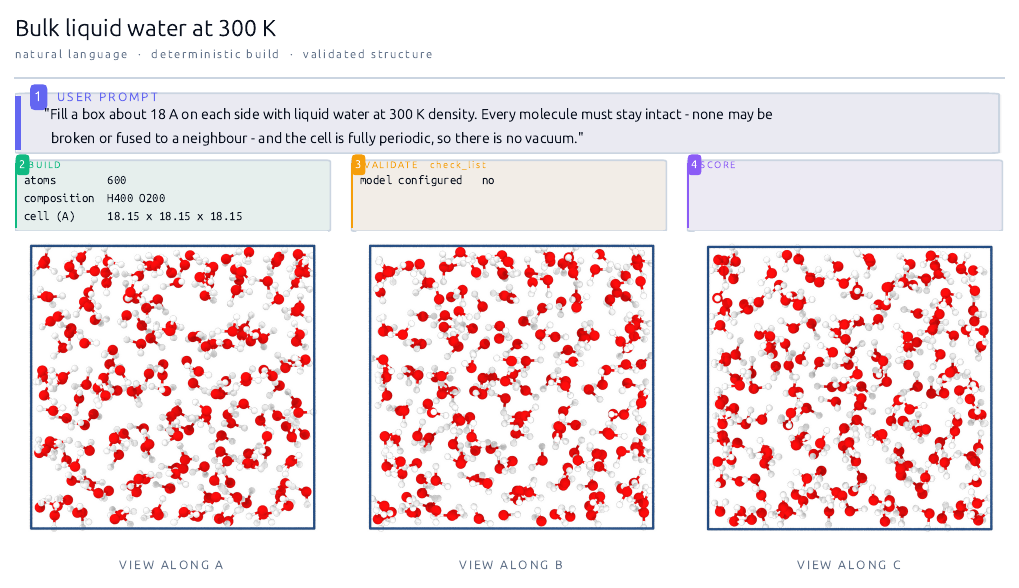}
\caption{Bulk molecular liquid. A cell of liquid water at 300~K density, generated by random sequential insertion of rigid molecules; every molecule is intact and the cell is fully periodic.}
\label{fig:case-bulk-water}
\end{figure}

\begin{figure}[t]
\centering
\includegraphics[width=\textwidth]{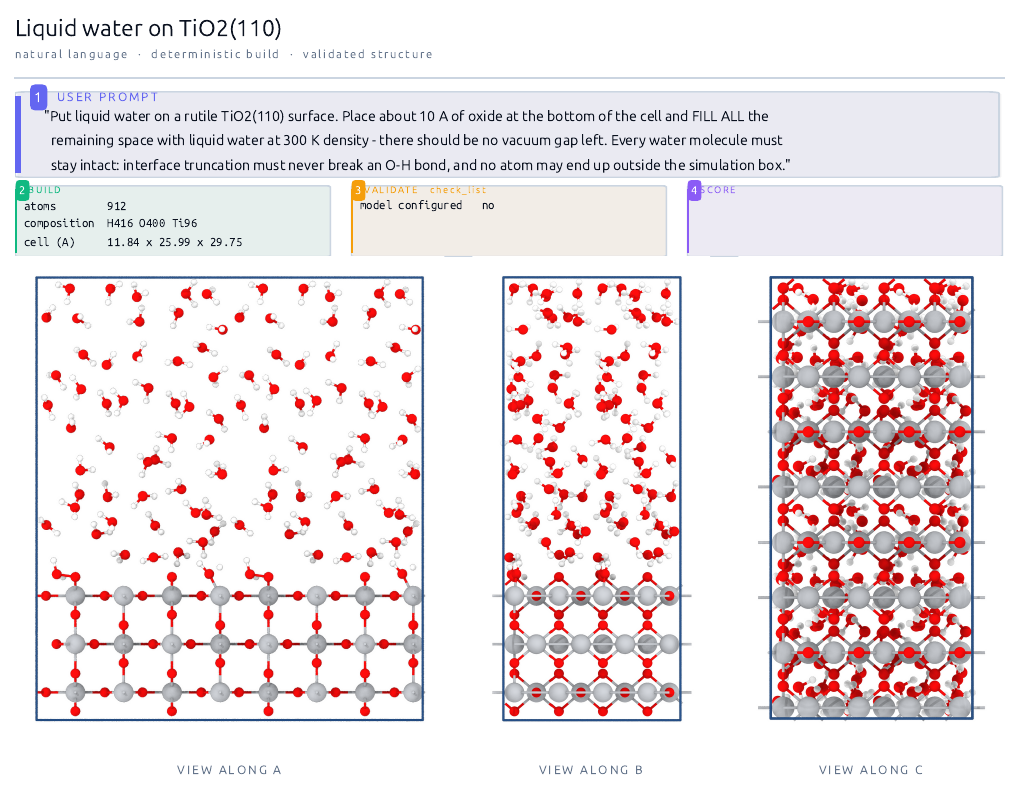}
\caption{Solid--liquid interface. Oxide at the bottom of the cell and liquid water filling everything above it, with no vacuum gap left. The interface truncation keeps whole molecules: an atom-wise cut would break O--H bonds.}
\label{fig:case-water}
\end{figure}

Each case figure presents one complete case---the natural-language request, the build, validation and score outcome, and the result along the three cell axes (Figures~\ref{fig:case-bulk}--\ref{fig:case-water}). A structure on its own does not show whether the pipeline understood the request: a wrong cell, a missing vacuum layer and a broken molecule all produce plausible-looking pictures. Putting the request and the outcome beside the result makes a failure attributable to a component rather than to the framework as a whole.

Two conventions are deliberate. Requests state \emph{physical sizes in \AA{}ngstr\"oms}---``about 16~\AA{} on each side'', ``a slab roughly 19~\AA{} thick'', ``a cavity about 12~\AA{} across''---rather than layer counts or $n\times n\times n$ replications. Someone asking for a structure does not know the lattice constant or the interlayer spacing of the material, and should not have to; the engine computes the replication that reaches the requested size (\S\ref{sec:c4}). And the simulation box is drawn on every view, including the 0D cases, where the box \emph{is} the model: it states how much vacuum surrounds the cluster, which is exactly the quantity a cluster calculation depends on.

Table~\ref{tab:cases} lists the fourteen builds; each is judged against its own request, and all fourteen are found consistent with it.

\begin{table}[h]
\centering
\caption{The fourteen demonstration builds, all produced by the engine from a specification. Every column is measured by the script that draws the figures (\texttt{make\_figures.py --table}), which builds each case and judges it against the prompt that asked for it with the scorer of \S\ref{sec:c6}: a build is scored 100 when the model reports no disagreement, and loses 25 points per major and 8 per minor disagreement. All fourteen are found consistent with their requests; the core--shell particle loses 8 points twice, on a 16.4~\AA{} core and a 3.2~\AA{} shell against the 16 and 3~\AA{} asked for.}
\label{tab:cases}
\begin{tabular}{@{}lrrll@{}}
\toprule
\textbf{Case} & \textbf{Atoms} & \textbf{Composition} & \textbf{Cell (\AA{})} & \textbf{Score} \\
\midrule
Bulk Si supercell & 216 & Si$_{216}$ & 16.3$^3$ & 100 \\
Au nanoparticle & 92 & Au$_{92}$ & 36.7$^3$ & 100 \\
Wulff Au nanoparticle & 807 & Au$_{807}$ & 40.5$^3$ & 100 \\
Au@Ag core--shell & 443 & Au$_{141}$Ag$_{302}$ & 32.4$^3$ & 84 \\
Si nanowire & 822 & Si$_{822}$ & 45.0$\times$48.9$\times$32.6 & 100 \\
Passivated Si cluster & 199 & Si$_{99}$H$_{100}$ & 21.7$^3$ & 100 \\
Si(111) slab & 432 & Si$_{432}$ & 23.0$\times$23.0$\times$31.6 & 100 \\
TiO$_2$(110) slab & 72 & Ti$_{24}$O$_{48}$ & 5.9$\times$13.0$\times$24.7 & 100 \\
Defect complex & 64 & Si$_{61}$P$_3$ & 10.9$^3$ & 100 \\
Si(111)$|$GaAs(111) & 1350 & Si$_{702}$Ga$_{324}$As$_{324}$ & 41.5$\times$41.5$\times$36.2 & 100 \\
C60 in bulk Si & 525 & Si$_{465}$C$_{60}$ & 21.7$^3$ & 100 \\
Amorphous SiO$_2$ & 648 & Si$_{216}$O$_{432}$ & 21.4$^3$ & 100 \\
Bulk liquid water & 600 & H$_{400}$O$_{200}$ & 18.1$^3$ & 100 \\
Water$|$TiO$_2$(110) & 912 & Ti$_{96}$O$_{400}$H$_{416}$ & 11.8$\times$26.0$\times$29.7 & 100 \\
\bottomrule
\end{tabular}
\end{table}

The set covers the dimensionality ladder: a periodic bulk cell (Figure~\ref{fig:case-bulk}); three kinds of 0D particle---a sphere cut (Figure~\ref{fig:case-nanoparticle}), a Wulff equilibrium shape (Figure~\ref{fig:case-wulff}) and a core--shell particle (Figure~\ref{fig:case-core-shell}); a 1D wire (Figure~\ref{fig:case-nanowire}); and a cluster passivated with hydrogen (Figure~\ref{fig:case-passivated}). Then the bounded and defected forms: a semiconductor slab (Figure~\ref{fig:case-slab}), an oxide slab (Figure~\ref{fig:case-tio2}) and a defect complex (Figure~\ref{fig:case-defect}). Finally the multi-component systems, where the difficulty lies in the interface rather than in either side: a lattice-matched heterojunction (Figure~\ref{fig:case-hetero}), a guest inside a carved cavity (Figure~\ref{fig:case-c60}), an amorphous network (Figure~\ref{fig:case-amorphous}), bulk liquid water (Figure~\ref{fig:case-bulk-water}) and a molecular liquid filling the cell above an oxide (Figure~\ref{fig:case-water}).

In every case the quantity the prompt asked for is visible in the figure---the vacuum, the gap between the two crystals, the cavity around the guest---because the box is drawn to scale. Beyond direct use, the same loop is a natural way to build training sets for machine-learning force fields. The classes above are the structural diversity such a set needs, a batch of them can be generated and validated without a person inspecting each one, and every structure carries a JSON specification, so the provenance of a training configuration is machine-readable; active-learning frameworks already automate the sampling and labelling side~\cite{scs}, but they take a valid starting structure for each system as given.


\section{Code Organization}
\label{sec:code}

The ATLAS codebase comprises 30 source modules---29 under \texttt{src/} plus the web app---totaling about 13,200 lines of Python, of which roughly a sixth is comments and docstrings that record why each rule exists. Alongside them sit the two vendored partner packages: a $\sim$6,300-line \texttt{cloud\_comp}~\cite{cloudcomp} that provides the primary analyzer layer, and a $\sim$2,300-line prototype classifier with its database~\cite{crystalproto}. A $\sim$640-line LLM skill file is the whole of C2's behaviour, and the figures are generated by a $\sim$1{,}900-line script rather than drawn. The code is built on NumPy~\cite{numpy}, SciPy~\cite{scipy}, scikit-learn~\cite{sklearn}, pymatgen~\cite{pymatgen} and Matplotlib, and organized by component and architectural layer (Table~\ref{tab:modules}); a 191-assertion suite over sixteen files and a 23-check regression harness pin the behaviour that the components promise each other.

\begin{table}[h]
\centering
\caption{Module Organization by Component}
\label{tab:modules}
\begin{tabular}{@{}l>{\raggedright\arraybackslash}p{8.2cm}r@{}}
\toprule
\textbf{Component} & \textbf{Modules} & \textbf{LOC} \\
\midrule
C1: pipeline analyzer & \texttt{analyzer} + \texttt{cloud\_comp} (partner package, $\sim$6,300 LOC), \texttt{utils}, \texttt{vendor} (which copy loads) & 556 ($+6{,}300$) \\
2D analysis lens & \texttt{projection\_engine}, \texttt{projection\_directions}, \texttt{density\_map}, \texttt{spatial\_features}, \texttt{cross\_view}, \texttt{asamp\_api} --- used by the interactive analysis views (C7), not by the build path & 1{,}659 \\
C2: LLM Skill          & \texttt{PROMPT\_TO\_JSON.md}, \texttt{llm\_client} (preview hook), \texttt{llm\_cache} (the cassette) & 1{,}513 \\
C3: JSON Architecture  & \texttt{component\_graph} (V4 algebra: components, operations, variables, check list), \texttt{paths} & 2{,}216 \\
C4: Build Engine       & \texttt{json\_executor} (entry point and source resolver), \texttt{measure}, \texttt{transform}, \texttt{strain}, \texttt{nanostructures}, \texttt{actions}, \texttt{merge}, \texttt{build\_flow} (orchestration) & 2{,}761 \\
C5: Validator          & \texttt{validate}                                               & 585 \\
C6: Scorer             & \texttt{scorer}, \texttt{materials} (formula + structure type = the material), \texttt{prompt\_check}, \texttt{chem\_text} (the shared chemistry reader) & 2{,}800 \\
C7: Web Interface      & \texttt{web\_prism}                                             & 1{,}043 \\
Figures                & \texttt{render\_ovito} (OVITO ray tracing; \texttt{render} is the matplotlib fallback), \texttt{figures/make\_figures} & 2{,}581 \\
\bottomrule
\end{tabular}
\end{table}

The crystals the engine builds from come from a bundled library of 17{,}773 CIF files (210~MB), organised by material class: 10{,}000 melt-quenched amorphous oxides, 7{,}461 fullerenes, and smaller sets of minerals, metal oxides, metals, ceramics, semiconductors and glasses, together with hand-built heterojunctions and a generated fullerene set. A \texttt{source} names a formula rather than a file path and matching is fuzzy, so a specification can ask for ``silicon'' without knowing which entry holds it, and the analyzer resolves the same name the same way. The library is located through \texttt{src/paths.py}, which prefers the copy inside the repository and accepts an environment override, so a build is reproducible from a fresh checkout.

The modules are layered shallowly, and the layers are there to be shared rather than called in sequence: \texttt{measure} (distances, coordination, the bond graph) and \texttt{transform} (geometric transforms) are the foundation; \texttt{nanostructures} and \texttt{strain} build the shapes and the lattice matches the specification vocabulary names; and \texttt{component\_graph} composes them into complete builds, driven by \texttt{build\_flow} in the web interface. Table~\ref{tab:modules} lists every module by component.

\section{Conclusion}
\label{sec:conclusion}

We have presented ATLAS, a framework that turns a written request into a validated atomic structure. Its central claim is that the two halves of the problem should be separated: a language model decides what is wanted, and a component algebra decides how to build it, with a JSON specification and a check list between them. That separation is what makes the result verifiable---a build is checked against the request rather than against its own assumptions, and a failure can be attributed to a named component---and it is what makes the pipeline scriptable end to end, from a sentence to a structure a calculation can use.

The separation turned out to buy something we did not set out to build. A boundary with a declared input and a declared output can be \emph{recorded}, and once the one stochastic stage is recorded the pipeline below it stops being a sequence of samples and becomes an instrument: the same request twice yields the same structure twice, a change to the code can be measured instead of guessed at, and a defect can be reproduced instead of argued about. It is also what lets a reader replay the fourteen demonstrations of \S\ref{sec:demo} without a model and without a network. The same property is why this loop is a plausible way to generate training data for machine-learning force fields: a set of structures whose provenance is machine-readable, whose validity has been checked against a stated request, and whose production is reproducible is exactly what such a set has to be.


\end{document}